\documentclass[reprint,amsmath,amssymb,aps,prl,groupedaddress,nofootinbib,twocolumn,superscriptaddress]{revtex4-1}
\usepackage{graphicx}
\usepackage{amsthm,amssymb,amsmath,braket,mathdots}
\usepackage{bm}
\usepackage[pagebackref=false,pdfnewwindow=true]{hyperref} 
\usepackage{epstopdf,psfrag}
\usepackage{relsize,amsbsy}
\usepackage[export]{adjustbox}
\usepackage{makecell}
\usepackage{graphicx,xcolor,tikz}
\usepackage{float}
\usepackage[centerlast]{caption}
\usepackage{hyperref}

\newcommand{\be}{\begin{equation}}
\newcommand{\ee}{\end{equation}}

\newcommand{\bit}{\begin{enumerate}}
	\newcommand{\eit}{\end{enumerate}}

\definecolor{bananayellow}{rgb}{1.0, 0.88, 0.21}
\definecolor{straw}{rgb}{0.32, 0.28, 0.1}

\begin{document}
	\title{Random multipolar driving: tunably slow heating through spectral engineering}
		\author{Hongzheng Zhao} 
		\author{Florian Mintert}
	\affiliation{\small Blackett Laboratory, Imperial College London, London SW7 2AZ, United Kingdom}
	\author{Roderich Moessner}
		\affiliation{\small Max-Planck-Institut f{\"u}r Physik komplexer Systeme, N{\"o}thnitzer Stra{\ss}e 38, 01187 Dresden, Germany}
	
		\author{Johannes Knolle  }
	\affiliation{Department of Physics TQM, Technische Universit{\"a}t M{\"u}nchen, James-Franck-Stra{\ss}e 1, D-85748 Garching, Germany}
	\affiliation{Munich Center for Quantum Science and Technology (MCQST), 80799 Munich, Germany}
	\affiliation{\small Blackett Laboratory, Imperial College London, London SW7 2AZ, United Kingdom}
	
	\begin{abstract}
 		Driven quantum systems may realize novel phenomena absent in static systems, but driving-induced heating can limit the time-scale on which these persist.
 		We study heating in interacting  quantum many-body systems driven by {\it random} sequences with  $n-$multipolar correlations, corresponding to a polynomially suppressed low frequency spectrum. For $n\geq1$, we find a prethermal regime, the lifetime of which grows algebraically with the driving rate, with exponent ${2n+1}$. A simple theory based on Fermi's golden rule accounts for this behaviour. The quasiperiodic Thue-Morse sequence corresponds to the $n\to \infty$ limit, 
 		and accordingly exhibits an exponentially long-lived prethermal regime. 
 		Despite the absence of periodicity in the drive, and in spite of its eventual heat death, the prethermal regime can host versatile non-equilibrium phases, which we illustrate with a random multipolar discrete time crystal.
	\end{abstract}

	\maketitle

{\it Introduction.--}
A closed quantum many-body system with 
a time-independent (static) Hamiltonian, by Noether's theorem, exhibits energy conservation. 
In thermodynamics, this underpins the notion of temperature, allowing for the distinction between high- and  low-temperature states, the former of which are typically disordered but the latter may display interesting correlations from symmetry breaking or topological order. 

By contrast, in systems with a time-dependent Hamiltonian, energy conservation is absent, as is the notion of (high or low) temperatures. Equilibrium states of driven systems are therefore entirely featureless ~\cite{d2014long,lazarides2014equilibrium,abanin2019colloquium}, often somewhat sloppily referred to as `infinite temperature' states. An initially ordered system, subject to a drive, thus equilibrates by absorbing energy until such a state with trivial correlations, i.e. described by a diagonal density matrix, is reached.

For periodically driven (`Floquet') systems, this heat death can be avoided by leaving the realm of equilibrium physics altogether: in disordered systems, a many-body localised (MBL) state is a non-ergodic alternative~\cite{abanin2019colloquium,ponte2015periodically}, stable to generic perturbations. This  may even host new forms of non-equilibrium order~\cite{moessner2017equilibration} -- such as spatiotemporal time-crystalline one~\cite{khemani_TC,else_TC},  without any counterparts in undriven systems. 

The observation that even in a clean system, before heating and equilibration take over,  there may be a  long-lived but transient `prethermal' regime has generated many interesting insights. For Floquet systems, which retain a notion of {\it discrete} time-translation symmetry, the existence of a prethermal regime is by now well-established. It is described by an effective Hamiltonian~\cite{kuwahara2016,Mori2016,Abanin2017,abanin2017rigorous,else2017prethermal,Rigol2019heating,Machado2020,machado2019,kuwahara2016,else2017prethermal,Takashi2019} derived in a high-frequency (Magnus) expansion, in which the driving period acts as a small parameter.   Similar prethermal regimes have also been found in static systems with only weakly broken conservation laws~\cite{Langen2016,bertini2015prethermalization,Mallayya2019,Luitz2020}.

A natural  question then is whether prethermalization can appear in driven systems even in the absence of the perfect periodicity of a Floquet system. With continuous quasiperiodic driving, such a possibility has been shown to exist, albeit under somewhat restrictive requirements on the analyticity of the drive~\cite{Else2020,Zhao2019,QPFlorian}. With discrete Fibonacci driving, a glassy relaxation has been identified~\cite{dumitrescu2018logarithmically}, but it cannot be captured by a prethermal Hamiltonian. For special integrable systems the slow down (or absence) of heating has also been reported in Ref.~\cite{nandy2017aperiodically,wen2020periodically,lapierre2020fine,crowley2019topological,korber2020interacting,nathan2019topological,boyers2020exploring,crowley2019halfinteger,peng2018time}.

\begin{figure}
	\centering
	\includegraphics[width=0.49\linewidth]{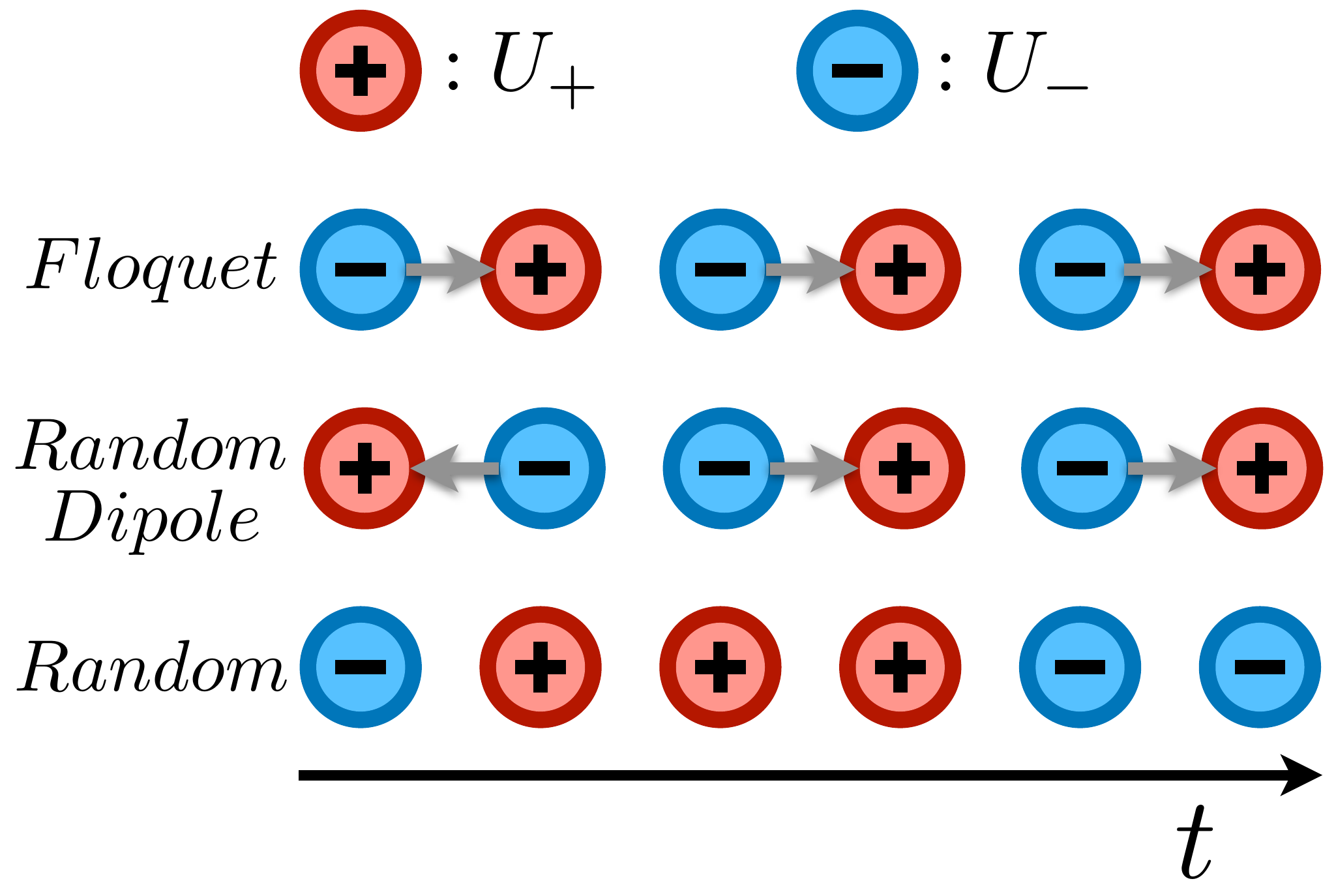}
	\includegraphics[width=0.49\linewidth]{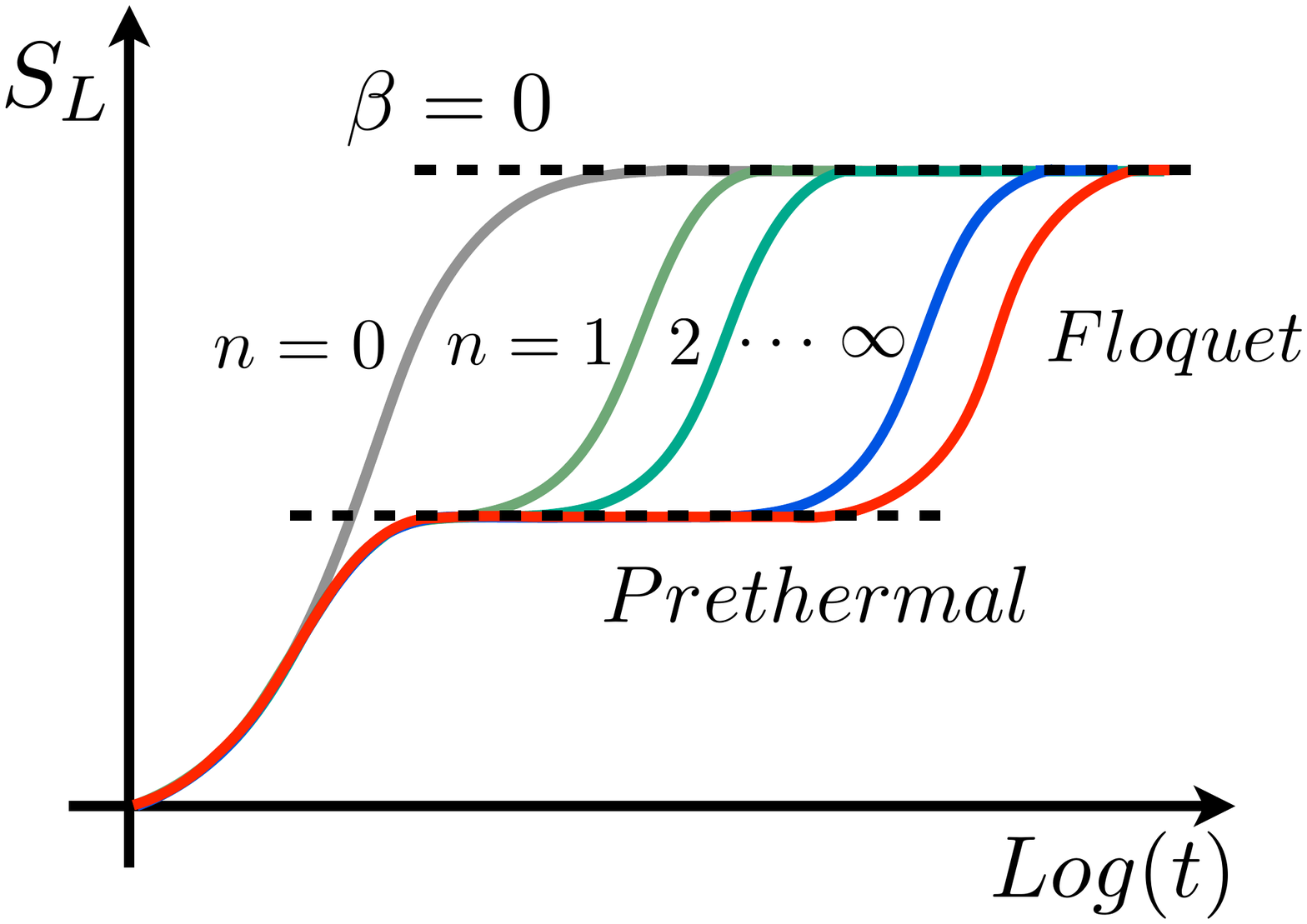}
	\caption{Left: schematic diagram for different driving protocols. Unitaries $U_\pm=\exp(-iT H_{\pm})$ can form two types of dipoles, $U_+U_-$ or $U_-U_+$. A Floquet drive is given by a perfect sequence of a single type of dipole whereas a random succession of both dipole types corresponds to a random dipolar drive. Right: Sketch of entanglement entropy dynamics. For a random drive $(n=0)$ the system heats rapidly to infinite temperature (inverse temperature $\beta=0$) but for $n-$multipolar ($n\geq1$) driving a prethermal plateau emerges. Quasiperiodic ($n\to \infty$) drives have an exponentially long (in driving rate) prethermal regime.}
	\label{fig:carton}
\end{figure}

Here we demonstrate that also {\it random} drives can lead to prethermal regimes described by effective Hamiltonians for generic quantum many-body systems.
To this end, we introduce a family of aperiodic but correlated drives that interpolate between highly structured quasiperiodic and fully random time dependence; the non-random versions correspond to Floquet drives, see Fig.~\ref{fig:carton}. It is based on random sequences of unitaries $U_\pm$, generated by two Hamiltonians $H_\pm$ acting for a time period $T$.
An integer $n$ parametrises the level of correlations incorporated in the sequence such that $n=0$ is a fully random sequence of $U_{\pm}$, while $n=1$ is made up of a random sequence of `dipoles', i.e.\ of terms $U_+U_-$ or $U_-U_+$; and in turn $n=2$, `quadrupolar' sequences, are made up of antialigned dipoles $U_-U_+U_+U_-$ or $U_+U_-U_-U_+$; and so on for higher $n$. The recursively defined limit $n\to\infty$ of such {\it random multipolar drives} (RMD) thus corresponds to the Thue-Morse quasiperiodic drive.

The basic motivation for considering such RMD is the observation that the bounds underpinning derivations of Floquet perthermalisation can be generalised to such settings, and do not per se require perfect Floquet periodicity. Our central finding is that for RMD energy absorption slows down {\it algebraically} with $n$, with the prethermal lifetime growing as $(1/T)^{2n+1}$. For the quasiperiodic $n\to\infty$ limit, this leads to an exponentially long prethermalisation scale. 

The correlated drives represent a form of spectral engineering as the Fourier transform of the  random time sequence of multipoles vanishes as a power $(1/T)^n$ yielding a gap in the limit $n\rightarrow\infty$. We find that this characteristic low frequency behavior underpins an ultimatley simple Fermi golden rule argument~\cite{Rigol2019heating,bilitewski2015scattering,Mallayya2019} accounting for the observed slow heating behavior and its $n$-dependence.

Finally, we show the existence of a prethermal random multipolar time-crystal, where a regular (im)perfect spin flip operation is sandwiched between random multipoles. Since the driving completely breaks time translation symmetry, this presents an unprecedented spatiotemporal phenomenon in many-body quantum systems~\cite{khemani2019brief}.

The remainder of this account is organised as follows. We first derive the rigorous bounds on heating for our random drives. To verify it numerically, we then define a generic model,  a clean driven non-integrable Ising model, Eq.~\ref{eq.model}, and present results of observables, counterposing numerical and analytical results for the prethermal lifetime. Finally, the presentation of the random multipolar time crystal precedes a concluding discussion.

{\it Rigorous bound.--}
\label{sec.scalingtheoretical}
We start from the two elementary time evolution operators as
\begin{eqnarray}
\begin{aligned}
\label{eq.A0}
U_+ = \exp(-iT H_+),\ \  U_- = \exp(-iT H_-),
\end{aligned}  
\end{eqnarray}
where $H_{\pm}$ is a time independent Hamiltonian and $T$ defines the characteristic time scale. For periodic driving, the dynamics is governed by the Floquet operator $U_1 = U_-U_+$, and the Floquet Hamiltonian $H_F$ is defined as $U_1=e^{-i2TH_{F}}$. It is well established how to construct a perturbative Floquet-Magnus (FM)  expansion in small $T$ (fast driving), $H_{F}=\sum_{n=0}^{\infty} (2T)^{n} H_F^n
$~\cite{kuwahara2016,Mori2016}.  

Already the zeroth order term, $H_{F}^0=(H_++H_-)/2$, is useful because of the following rigorous bound for the error accumulated over a single period~\cite{kuwahara2016floquet}:
 \begin{equation}
\label{eq.lowestorder}
\left\|U_-U_+-e^{-i H_{F}^0 2T}\right\| \leqslant V_0 \left[6\cdot 2^{-n_{0}}+\lambda T\right]2T,
\end{equation} 
where  $V_0$ is proportional to the driving amplitude, $\lambda$ captures the local typical energy scale of the system, and $n_{0}\sim \mathcal{O}(T^{-1})$ denotes the optimal order before the FM expansion diverges~\cite{kuwahara2016floquet,appendix}. 
 Note that Eq.~\ref{eq.lowestorder} does not depend on the order of $U_+$ and $U_-$, because the paired operators $U_+U_-$ and $U_-U_+$ have the same $H_F^0$. Consequently, using a triangle inequality, the error for the time evolution up to $t=2mT$ can be estimated to be
 \begin{equation}
\label{eq.scaling}
\Big \| \underbrace{{(U_-U_+)(U_+U_-)\dots}}_\text{$m$ cells} -e^{-i H_{F}^0 t}\Big\|
\leqslant V_0 \left[6\cdot2^{-n_{0}}+{\lambda T}\right]t.
\end{equation}
Crucially, the bound is indeed not limited to periodic driving and the paired operators can appear in random sequence. Despite the fact that the derived bound is not tight, Eq. \ref{eq.scaling} indicates that in the fast driving regime the error only becomes notable after a sufficiently long time leading to a long-lived prethermal regime~\cite{kuwahara2016floquet} whose dynamics can be approximated by the effective Hamiltonian $H_{F}^0$ with a quasi-conserved energy~\cite{Rigol2008}. 

In the following, we confirm via exact diagonalization (ED) the possibility of prethermal regimes as indicated by the bound, the lifetime of which will be further justified via a Fermi's golden rule calculation.

{\it Prethermalization.--}
We focus on a generic spin model described by the Hamiltonian
\begin{equation}
H_{\pm} = \sum_i J_x \sigma_i^x\sigma_{i+1}^x +J_z\sigma_i^z\sigma_{i+1}^z +(B_0\pm B_x)\sigma_i^x+B_z\sigma_i^z,
\label{eq.model}
\end{equation}
 where $J_x, J_z$ are the nearest-neighbor exchange interactions, $B_0,B_z$ are static fields, and $B_x$ denotes the driving amplitude. We use periodic boundary conditions such that translation invariance allows us to access the  dynamics for larger system sizes. 
 
 To characterize the thermalization dynamics, we use two different diagnostics. First, we calculate the mean energy of the effective system $\langle H_{F}^0\rangle$, which should remain constant in the prethermal regime but drop to zero once the state heats up to infinite temperature. Second, we study the growth of the half-chain entanglement entropy in time $S_{L}(t) =-\operatorname{Tr}\left[\rho_{L/2} \log \rho_{L/2}\right]$ with the half chain reduced density matrix $\rho_{L/2}=\operatorname{Tr}_{1\leq i \leq L/2}\left\lbrace\ket{\psi(t)}\langle \psi(t)|\right\rbrace$.

We first consider the quasiperiodic Thue-Morse driving~\cite{nandy2017aperiodically}, the
$n\to \infty$ limit of $n-$RMD. Starting from Eq.~\ref{eq.A0} we use $U_1 = U_-U_+$ and $\tilde U_1 = U_+U_-$ to recursively construct the driving unit cells of time length $2^nT$ as 
\begin{eqnarray}
\begin{aligned}
U_{n+1} = \tilde U_n U_n, \  \  \tilde U_{n+1}  =U_n \tilde U_n
\end{aligned}
\label{eq.definition}
\end{eqnarray} Note, $n-$RMD random sequences are generated from unit cells $U_n$ and $\tilde U_n$.
This recursive construction enables us to simulate the dynamics for exponentially long time only involving a linearly increasing number of matrix multiplications, i.e. $\ket{\psi(2^nT)}=U_n\ket{\psi(0)}$.

\begin{figure}
	\centering
	\includegraphics[width=1\linewidth]{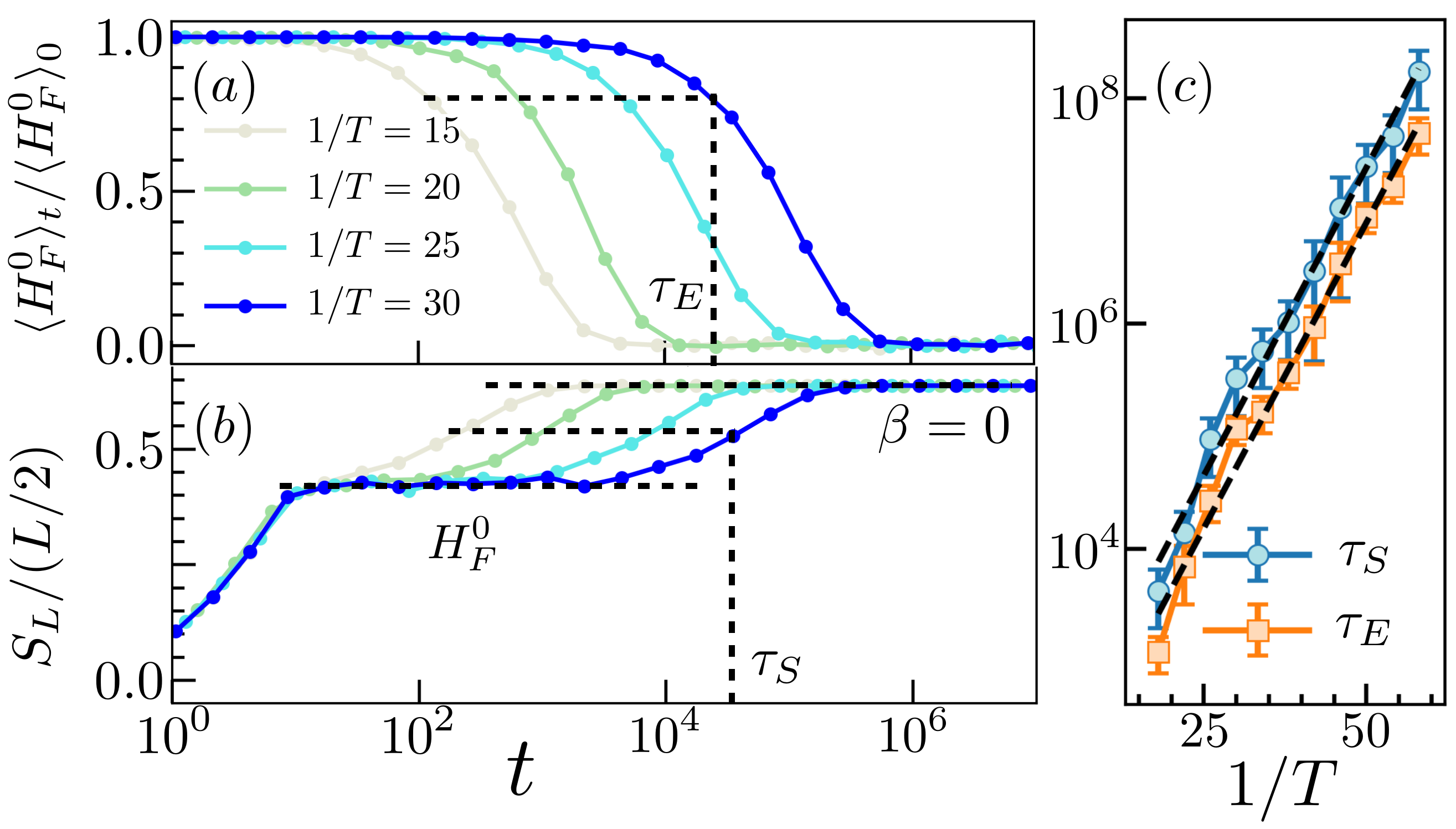}
	\caption{Evolution of (a) mean energy  and (b) entanglement entropy  with Thue-Morse driving as a function of characteristic time $T$, using parameters $\{J_z,J_x,B_x,B_z,B_0\}=\{1,0.243,0.809,0.357,0.21\}$, $L=18$. Prethermalization appears with an entanglement plateau and a lack of energy absorption. (c) Exponential scaling of the thermalization time versus $1/T$.}
	\label{fig:dynamics}
\end{figure}

In  Fig.~\ref{fig:dynamics} the dynamics of the mean energy density and the entanglement entropy are shown for different driving rates $1/T$ at stroboscopic time $2^n T$. The initial state is taken with all spins pointing down. For sufficiently fast driving $1/T\geq20$, the entanglement entropy $S_L$ first saturates to a prethermal plateau, which is well captured by the zeroth-order effective Hamiltonian. Meanwhile, the system heats only very slowly from the external drive, hence, the mean energy remains constant over a large time window. In between stroboscopic times, we also verify that the mean energy remains quasi-conserved. Only after a large time scale $\tau_S$, the entropy $S_L$ rapidly grows to the final plateau with $S^{\beta=0}_L=[L\ln(2)-1]/2$~\cite{page1993average}, which confirms that the system has eventually thermalized to the infinite temperature state. Correspondingly, the mean energy drops to zero after the time scale $\tau_E$. 

Next, we study the scaling of the thermalization times as a function of frequency. $\tau_{E,S}$ can be extracted by setting certain thresholds, e.g. $S_L/(L/2)\sim 0.55\pm0.05$ or $\langle H_{F}^0\rangle_t/\langle H_{F}^0\rangle_0\sim 0.7\pm0.1$. In Fig.~\ref{fig:dynamics}(c)  we show that both definitions of the thermalization have the same exponential growth with frequency. We have verified that the scaling does not show qualitative dependence on the initial states or precise values of the threshold values~\cite{appendix}.  

\begin{figure}
	\centering
	\includegraphics[width=0.92\linewidth]{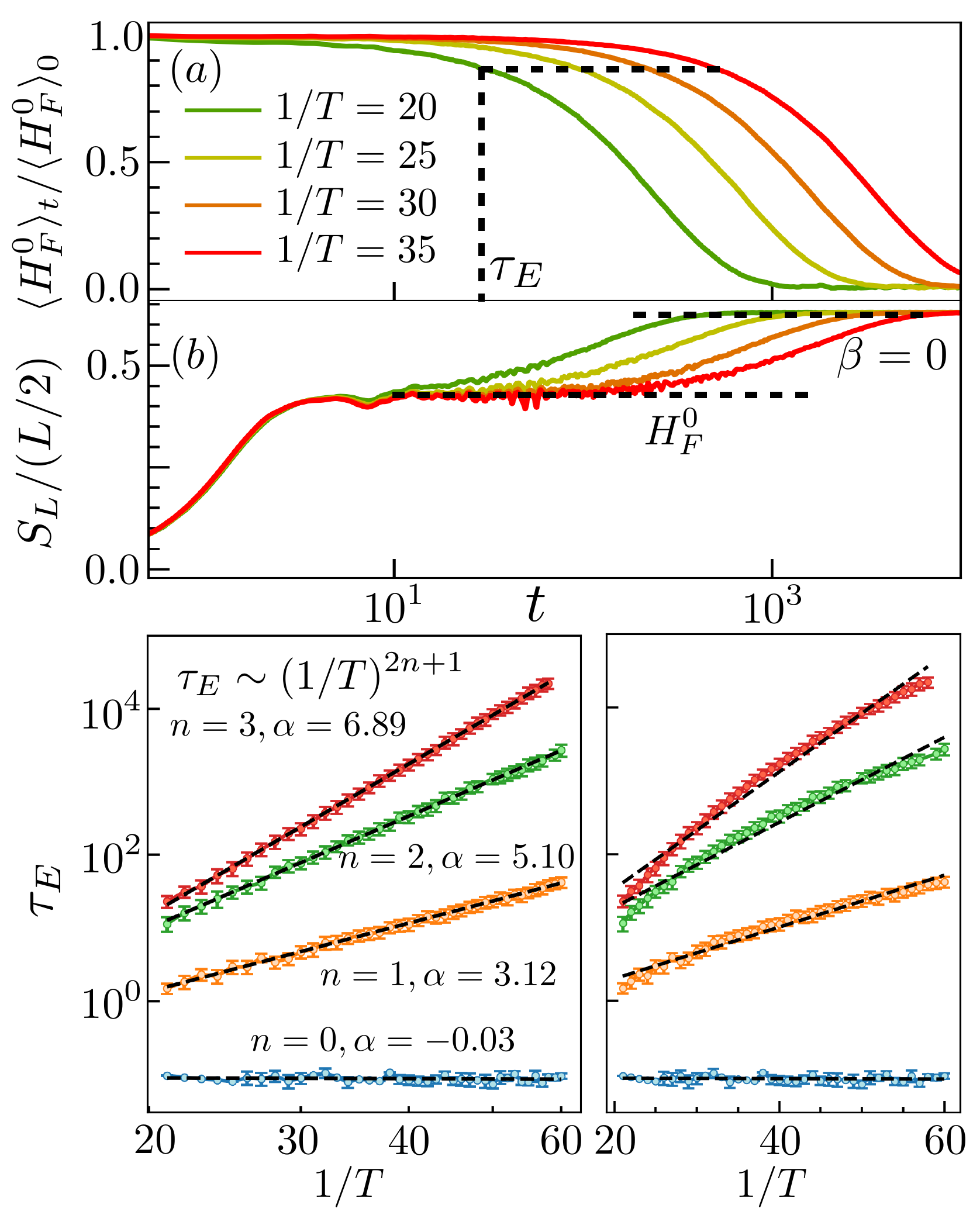}

	\caption{Evolution of (a) mean energy and (b) entanglement entropy with quadrupolar random driving, with parameters $\{J_z,J_x,B_x,B_z,B_0\}=\{1,0.71,3.2,0.25,0.21\}$, $L=16$. Prethermal plateau of entanglement entropy appears for $1/T\geq 20$. Thermalization time dependence on $1/T$ in log-log (c) and log  scale (d) for $n-$multipolar drivings which confirms the algebraic dependence $(1/T)^{\alpha}$ with $\alpha\approx 2n+1$ for $n\geq 1$}. 
	\label{fig:multipole}
\end{figure}

We verified that the prethermal regime also exists for RMD with finite $n$ and not only for the  Thue-Morse limit. The upper two panels of Fig.~\ref{fig:multipole} depict the results for a random quadrupolar drive. A similar prethermal plateau can be identified from the entanglement entropy (Fig.~\ref{fig:multipole} (b)) and the lack of energy absorption (Fig.~\ref{fig:multipole} (a)). As before, we determine thermalization time by the thresholds $\langle H_{F}^0\rangle_t/\langle H_{F}^0\rangle_0\sim 0.96\pm0.01$. 

In Fig.~\ref{fig:multipole} (c), we show the scaling of $\tau_E$ on a double log plot for different RMD structures with $n=0,1,2,3$. In contrast to the exponential scaling observed for the Thue-Morse driving, we identify an algebraic dependence 
\begin{eqnarray}
\tau_E \propto (1/T)^{\alpha} \   \  \text{with} \   \ \alpha \approx 2n+1,
\label{eq.thermtime}
\end{eqnarray}
for $n\geq1$.
Interestingly, the fitted exponent strongly depends on the multipolar correlations and is to a good accuracy a simple function of $n$. We have verified that the scaling exponent is robust to change of parameters. As a comparison, we also plot the same result on a log scale in Fig.~\ref{fig:multipole} (d) which indicates a clear deviation from an exponential fit towards larger $1/T$, especially for $n\geq 2$.  

The thermalization time of the purely random drive $n=0$ is short ($\tau_E<1$) and independent on the driving rate, which demonstrates a qualitative improvement for heating suppression by the multipolar structure. Note, we also identify that random drives $n=0$ may follow Eq.\ref{eq.thermtime} with exponent $\alpha=1$, see Supp. Mat.~\cite{appendix}, but the crucial difference to the case of $n>0$ is that such scaling is only observed in the perturbative regime for small driving amplitudes.

{\it Fermi's golden rule.--} Although the bound in Eq.~\ref{eq.scaling} indicates the existence of a prethermal regime, it is not tight and insufficient to predict the scaling of the thermalization time. As an alternative, we show in the following that the characteristic scaling follows from a simple extension of Fermi's golden rule (FGR) as applied to Floquet systems~\cite{bilitewski2015scattering,Bilitewski2015,Rigol2019heating,weidinger2017,lellouch2017parametric}. 

Let us first follow Ref.~\cite{Rigol2019heating} to recall the FGR for periodically driven systems described by the Hamiltonian $\hat{H}(t)=\hat{H}_{0}+ g(t) \hat{K},$ where $g(t) \hat{K}$ denotes the weak periodic driving with $g(t)=\sum_m g_m\sin(m\Omega t)$ with $\Omega=2\pi/T$ and $g_m$ the strength of the Fourier components. The thermalization rate can be written as $ \Gamma(t)=\sum_m \dot{E}_m(t) /\left[E_{\infty}-E(t)\right]$, where $\dot{E}_m$ denotes the average rate of energy absorption for mode $m$ and $E_{\infty}$ is the energy of the system at infinite temperature. In the linear response regime, this rate remains almost constant and its inverse enables one to estimate the thermalization time scale~\cite{Rigol2019heating}. For fast drivings, extensive numerical evidence~\cite{Rigol2019heating,machado2019} and theoretical analysis~\cite{abanin2015exponentially,ho2018bounds,Mori2016,Abanin2017,kuwahara2016,else2017prethermal,Tran2019} suggest that the thermalization rate is exponentially suppressed as
$
\Gamma=\sum_m g_m^2 Ae^{-m\Omega/\epsilon},
$
where $A,\epsilon$ are both model dependent parameters. Accordingly, the prethermal regime is exponentially long-lived as $\tau \propto e^{\Omega}$ for Floquet systems.

Next, we can extend the FGR to the $n-$RMD with a continuous frequency spectrum 
$g(t)=\int dx g_x \sin (x \Omega t)$
with an algebraically suppressed weight at low frequencies $g_x\propto x^n$ as follows from the auto-correlation function of the multipolar sequence generated from Eq.~\ref{eq.definition}, see Supp. Mat.~\cite{appendix}. Again, in the linear response regime it is assumed that the system absorbs energy from each frequency mode independently such that  
\begin{eqnarray}
\Gamma \propto \int_0^\infty \text{d}x x^{2n} Ae^{-x\Omega/\epsilon} \propto \Omega^{-(2n+1)}.
\label{eq:heating1}
\end{eqnarray}
Correspondingly, the thermalization time scales as $(1/T)^{2n+1}$ in accordance with the numerics,  Fig.~\ref{fig:multipole}(c).

The Fourier spectrum of the quasiperiodic TMS driving vanishes as $x^{(n\to \infty)}$ for $x\to 0$, effectively generating a gap proportional to $\Omega$, see Supp. Mat.~\cite{appendix}. Therefore, the most dominating heating rate is given by the smallest allowed frequency and we can simply model $g_x\propto \delta(x-x_0)$ to obtain
\begin{equation}
\Gamma_{TMS} \propto \int_{0}^{\infty} \mathrm{d} x \delta(x-x_0)  A e^{-x \Omega / \epsilon} \propto e^{-x_0 \Omega / \epsilon},
\end{equation} with $x_0\Omega$ the gap size. The heating process hence becomes similar to that of Floquet systems: if the gap is larger than the local band width $J$, multiple spin-flips involving at least $x_0\Omega/J$ spins are needed to absorb energy from the drive~\cite{abanin2015exponentially,Mori2016}, but as theses collective processes are rare, it leads to an exponential scaling of the thermalization time $ \mathcal{O}(e^{1/T})$ in accordance with the numerical results of Fig.~\ref{fig:dynamics}.

Overall, the agreement of the numerical results and the FGR rate Eq.~\ref{eq:heating1} leads to a surprisingly simple picture, namely, that the dominant heating is induced by the absorption of single low energy modes even for the continuos spectrum, whereas the inevitably present multi-mode processes only contribute at later time scales when the system has already thermalized. 

{\it Prethermal Random Multipolar DTC.--}
\begin{figure}
	\centering
	\includegraphics[width=1\linewidth]{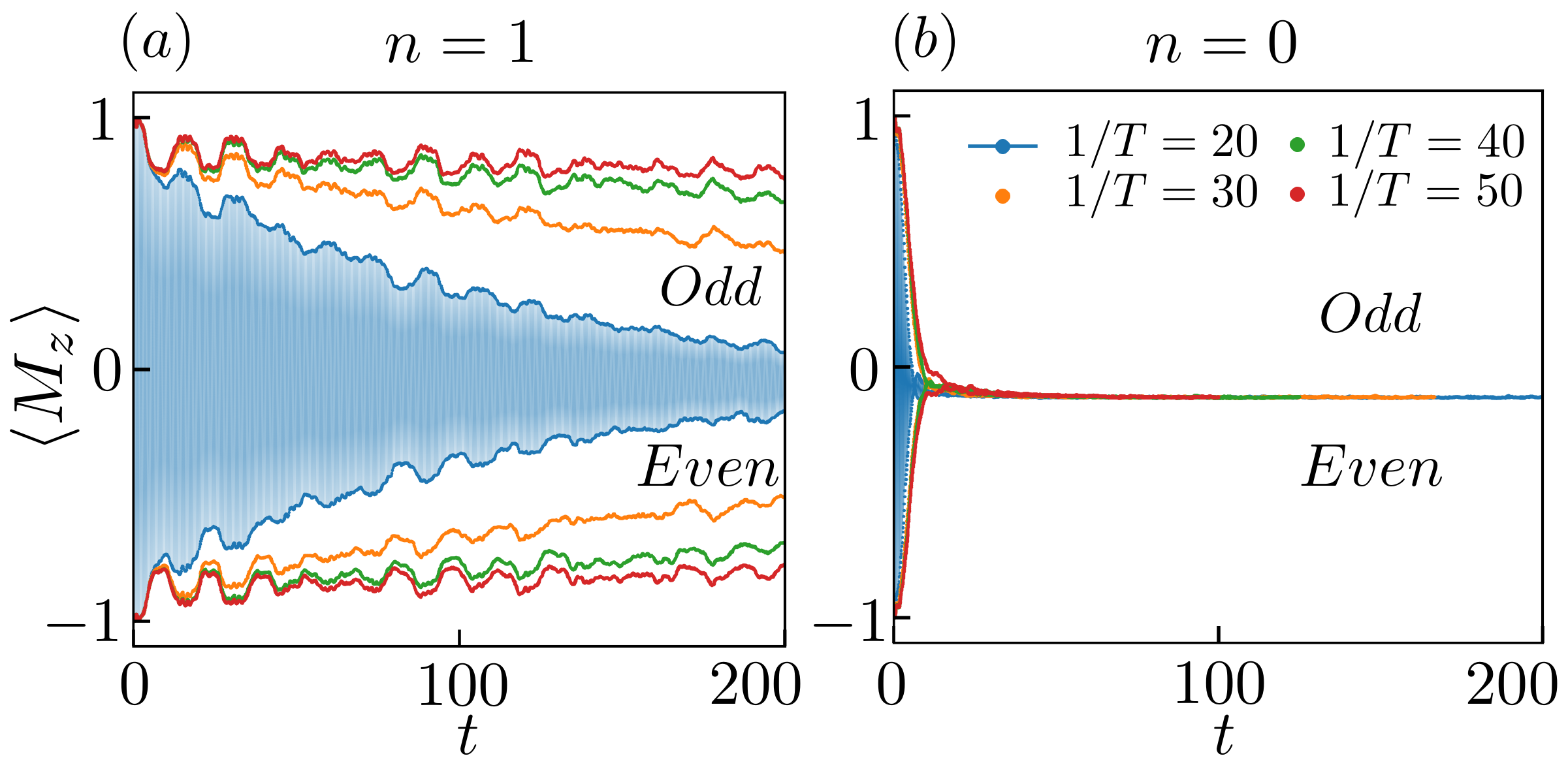}
	\caption{Dynamics of the magnetization of the central spin induced by a random dipolar drive with additional spin flips, using parameters $\{J_z,J_x,B_z,B_0,B_x\}= \{1,0.315,0.21,-0.05,0.75\}, L=18$. A long-lived prethermal DTC exists with $n=1$ RMD driving (a), while heating is inevitably fast for the $n=0$ random case (b).  
	}
	\label{fig:randomdtc}
\end{figure}
Finally, we provide a concrete example of a prethermal non-equilibrium phase for our family of RMDs. We extend the idea of Floquet DTCs~\cite{khemani_TC,else_TC}
to a situation where the drive contains {\it temporally random} components between the spin flips: for example,
we add global spin flips in between the dipolar time evolution operators as
\begin{eqnarray}
\begin{aligned}
\label{eq.DTC}
U_1' = e^{-iH_-T}e^{-iH_+T}X, \ \tilde U_1' = e^{-iH_+T}e^{-iH_-T}X,
\end{aligned}
\end{eqnarray}
where 
$X = \exp \left(i \frac{\pi}{2} \sum_{i} \sigma_{i}^{x}\right)
\sim \prod_{i} \sigma_{i}^{x}$. 
According to our discussion above, both dipolar operators can be approximated as
\begin{eqnarray}
U_1', \tilde U_1'\approx e^{-i(H_++H_-)T}X,
\end{eqnarray}
to lowest order in the Magnus expansion.

For the ideal case when the effective Hamiltonian, $H_{F}^0=(H_++H_-)/2$, preserves the $\mathbb{Z}_2$ Ising symmetry, products of $U_1', \tilde U_1'$ can be approximated as $(e^{-iH_F^02T}X)^2=e^{-iH_{F}^04T}$ such that in the prethermal regime the time evolution at stroboscopic times $4T$ is well described by the effective Hamitonian $H_{F}^0$. Consequently, for a $\mathbb{Z}_2$ symmetry broken initial state, the local magnetization will exhibit period-doubling behavior with respect to the $2T$ periodic spin flips. Such a random prethermal DTC should still persist if the $\mathbb{Z}_2$ symmetry of $H_F^0$ is weakly broken, but the lifetime will decrease depending on the perturbation. 

In Fig.~\ref{fig:randomdtc} (left panel), we have  numerically verified the existence of a prethermal DTC with random dipolar driving. The lifetime notably  increases for faster driving, in particular for $T^{-1}=50$ the amplitude of the period-doubling magnetization does not decrease appreciatively for numerically accessible times. We also verified the robustnes of the prethermal DTC for imperfect spin flip operations~\cite{appendix}. Fig.~\ref{fig:randomdtc} contrasts the prethermal nature of RMD to the one of a purely random drive built from
\begin{eqnarray}
\begin{aligned}
U_+' = e^{-iH_+T}X, \ U_-' = e^{-iH_-T}X.
\end{aligned}
\end{eqnarray}
As $U_+',U_-'$ also perform perfect spin flips, a period-doubling pattern exists for short times $t\lesssim10$, but heating is inevitably fast and no prethermal regime appears.  

{\it Conclusions and outlook.--}
We have introduced a family of {\it random} drives with $n$-multipolar correlations in time which give rise to prethermal regimes in interacting many-body quantum systems. We found numerically, and via a Fermi's golden rule calculation, a characteristic algebraic dependence of the heating time scale on $n$. The quasiperiodic limit $n\to \infty$ of the Thue-Morse sequence displays an exponentially long-lived prethermal regime.

Random multipolar drives present an elementary and controlled way to introduce randomness to a Floquet system. The resulting tunable algebraic prethermalisation adds a new aspect 
to our understanding of paths towards thermalization. Beyond Fermi's golden rule, it remains intriguing how multi-photon processes~\cite{weinberg2015multiphoton} contribute to the late-stage heating. With numerics limited to small systems, as is often the case for generic interacting many-body systems, more extensive numerical studies would also be worthwhile. 

Beyond Floquet DTCs~\cite{khemani_TC,else_TC} or discrete time quasi-crystals~\cite{Zhao2019,dumitrescu2018logarithmically}, the random drives completely break discrete time translation symmetry, and thus enrich the growing zoo of non-equilibrium phases of matter. 

RMD represents a simple form of spectral engineering in driven systems. It would be interesting to study this in relation to  many-body localization and eigenstate order~\cite{DTC,dumitrescu2018logarithmically}, or in the context of quantum information processing. Finally, an obvious much broader question concerns the scope of such 
spectral engineering in non-equilibrium quantum many-body dynamics more generally. 

{\it Acknowledgements.--}
We acknowledge helpful discussion with Takashi Mori, Hyukjoon Kwon and Joseph Vovrosh. The work was in part supported by the Deutsche Forschungsgemeinschaft  under grants SFB 1143 (project-id 247310070) and the cluster of excellence ct.qmat (EXC 2147, project-id 390858490), and by a fellowship within the Doctoral-Program of the German Academic Exchange Service (DAAD). We acknowledge support from the Imperial-TUM flagship partnership.

\newpage
\bibliography{Prethermal}
\bibliographystyle{unsrt}
\appendix

\section{Proof of the bound Eq.~\ref{eq.scaling}}
Here closely following Ref.\cite{kuwahara2016floquet} we derive the bound for the time evolution operators of the random dipolar drives, which is made up of random sequences of dipoles $U_1=U_-U_+$ or $\tilde U_1=U_+U_-$ with 
\begin{eqnarray}
\begin{aligned}
U_+ = \exp(-iT H_+), U_- = \exp(-iT H_-).
\end{aligned}  
\end{eqnarray}
The time-dependent Hamiltonian for each dipole can be written respectively as
\begin{eqnarray}
H^{A}(t) &=& H_- (0<t<T), H_+(T<t<2T), \\
H^{B}(t) &=& H_+ (0<t<T), H_-(T<t<2T), 
\end{eqnarray}
which can also be rewritten as
$H^{A/B}(t)= H_{\mathrm{static}}+V^{A/B}(t)$, with $H_{\mathrm{static}}$ being the time-independent part. We consider a general few body Hamiltonian involving at most k-body interactions with finite k:
\begin{equation}
H_{\mathrm{static}}=\sum_{|X| \leq k} h_{X}, \quad V^{A/B}(t)=\sum_{|X| \leq k} v^{A/B}_{X}(t),
\end{equation}
where $X$ labels the set of interacting sites and $|X|$ is the size of the set. Note, we also have the property
\begin{eqnarray}
\begin{aligned}
\label{eq.same}
v_X^{A/B}(0<t<T) = v_X^{B/A}(T<t<2T),
\end{aligned}
\end{eqnarray}  which will be used later.

We use the parameter $J^{B/A}$ to denote the local interaction strength (or single particle energy) of the system: 
\begin{equation}
\sum_{X: X \ni i}\left(\left\|h_{X}\right\|+\|v^{A/B}_{X}(t)\|\right) \leq J^{A/B},
\end{equation}
where $||\dots||$ is the operator norm, and $\sum_{X: X \ni i}$ denotes the summation w.r.t. the supports containing the spin $i$. Based on Eq.~\ref{eq.same}, we can easily see $J^{A}=J^{B}:=J$ and define $ \lambda:=2 k J$ as the typical local energies of the system.  
We introduce the average driving norm as
\begin{eqnarray}
\begin{aligned}
V^{A/B}_{0}:=\sum_{|X| \leq k} \frac{1}{2T} \int_{0}^{2T}\left\|v^{A/B}_{X}(t)\right\| d t,
\end{aligned}
\end{eqnarray}
where $2T$ is used because each dipole takes time $2T$. Moreover by separating the time integral and using Eq.~\ref{eq.same}, one arrives at
\begin{eqnarray}
\begin{aligned}
\label{eq.summation}
&V^{A/B}_{0}=\sum_{|X| \leq k} \frac{1}{2T} \Big[\int_{0}^{T}\left\|v^{A/B}_{X}(t)\right\| d t
+\int_{T}^{2T}\left\|v^{A/B}_{X}(t)\right\| d t\Big]\ \ 
\\&=\sum_{|X| \leq k} \frac{1}{2T} \int_{0}^{T}\Big[\left\|v^{B/A}_{X}(t)\right\| 
+\left\|v^{A/B}_{X}(t)\right\| \Big]d t.
\end{aligned}
\end{eqnarray}
One can realize that  $V_0^{A}=V_0^{B}$, which will be denoted as $V_0$ in the following. 

According to Ref.\cite{kuwahara2016floquet}, both of the dipole operators, $U_1$ and $\tilde U_1$, can be approximated by the same zeroth order effective Hamiltonian $H_F^{0}=(H_++H_-)/2$, and the error is bounded as
\begin{eqnarray}
\label{eq.singleperioid}
\left\|U_1/\tilde U_1-e^{-i H_{F}^0 2T}\right\| \leqslant [6\cdot 2^{-n_{0}}+\lambda T]V_02T.
\end{eqnarray}
The exponent \begin{equation}
n_{0}:=\left\lfloor\frac{1}{16 \lambda (2T)}\right\rfloor=\mathcal{O}(T^{-1}),
\end{equation} with the floor function $\lfloor.\rfloor$, denotes the optimal order of the Floquet-Magnus expansion before it diverges~\cite{kuwahara2016floquet}. The error accumulates during the time evolution hence the dynamics deviate from the one $H_F^0$ predicts. The rigorous bound in Eq.~\ref{eq.scaling} can be obtained by using Eq.~\ref{eq.singleperioid} and the triangle inequality for arbitrary unitaries
\begin{eqnarray}
\begin{aligned}
&\left\|W_1W_2-V_1V_2\right\|=\left\|(W_1-V_1)W_2+V_1(W_2-V_2)\right\|\\ &\leq\left\|(W_1-V_1)W_2\right\|+\left\|V_1(W_2-V_2)\right\|\\
&= \left\|W_1-V_1\right\|+\left\|W_2-V_2\right\|,
\end{aligned}
\end{eqnarray}
where we used that unitary operators do not change the norm in the last equality. 

\paragraph{Improvement of the bound for $n\geq2$ }
For $n\geq 2$, one can improve the bound for the time evolution operators  \begin{eqnarray}
\begin{aligned}
U_2 &= \tilde{U}_1U_1=U_+U_-U_-U_+,\\\tilde{U}_2 &= U_1\tilde{U}_1=U_-U_+U_+U_-,
\end{aligned}
\end{eqnarray}
by realizing the first order contribution $\mathcal{O}({T})$ to the magnus expansion for $U_2/\tilde{U}_2$ vanishes due to the above symmetric construction of quadrupoles. Therefore, both the operators can be well-approximated through the lowest order effective Hamiltonian $H_F^0$ as~\cite{kuwahara2016floquet}
\begin{equation}
\left\|U_2/\tilde{U}_2-e^{-i  H_{F}^{0} 4T}\right\| \leqslant \left[6 \cdot 2^{-n'_{0}}+\frac{4}{9}(4 T \lambda)^{2}\right] V_{0}4 T,
\end{equation}
with the exponent 
$
n'_{0}:=\left\lfloor{1}/{16 \lambda (4T)}\right\rfloor=\mathcal{O}(T^{-1}).
$ Hence, the error for the time evolution up to $t=4mT$ can be estimated to be
\begin{equation}
\label{eq.scaling}
\Big\|\underbrace{{U_2\tilde{U}_2\dots}}_\text{$m$ quadrupoles} -e^{-i H_{F}^0 t}\Big\|
\leqslant V_0 \left[6\cdot2^{-n'_{0}}+\frac{4}{9}(4 T \lambda)^{2}\right]t.
\end{equation}

\section{Algebraic suppression of the frequency spectrum}
In this section, we discuss the Fourier spectrum for random multipolar sequences. We will first show numerical results about the scaling of the suppression at low frequency, then rationalize it analytically via the autocorrelation function.

In contrast to the $n-$random multipolar drives constructed from unitaries $U_-,U_+$, here we replace unitaries by integers $-1,+1$ for generating the random multipolar sequence in time. For instance for $n=0$, 
we have the random sequence  $x^{(0)}(t)=\{1,1,-1,1,-1,\dots,+1\}$; and the $n=1$ sequence is made up of random dipolar unit cells $(-1,+1),(+1,-1)$; for $n=2$, two anti-aligned dipoles form quadrupolar unit cells as $(-1,+1,+1,-1)$ and $(+1,-1,-1,+1)$.
Note, these two unit cells differ by a relative `-1' sign, which will be used to determine the recursive relation for the autocorrelation function.

We now can compare the real part of the discrete Fourier transformation
of sequences for different $n$.
\begin{figure}[h]
	\centering
	\includegraphics[width=0.47\linewidth]{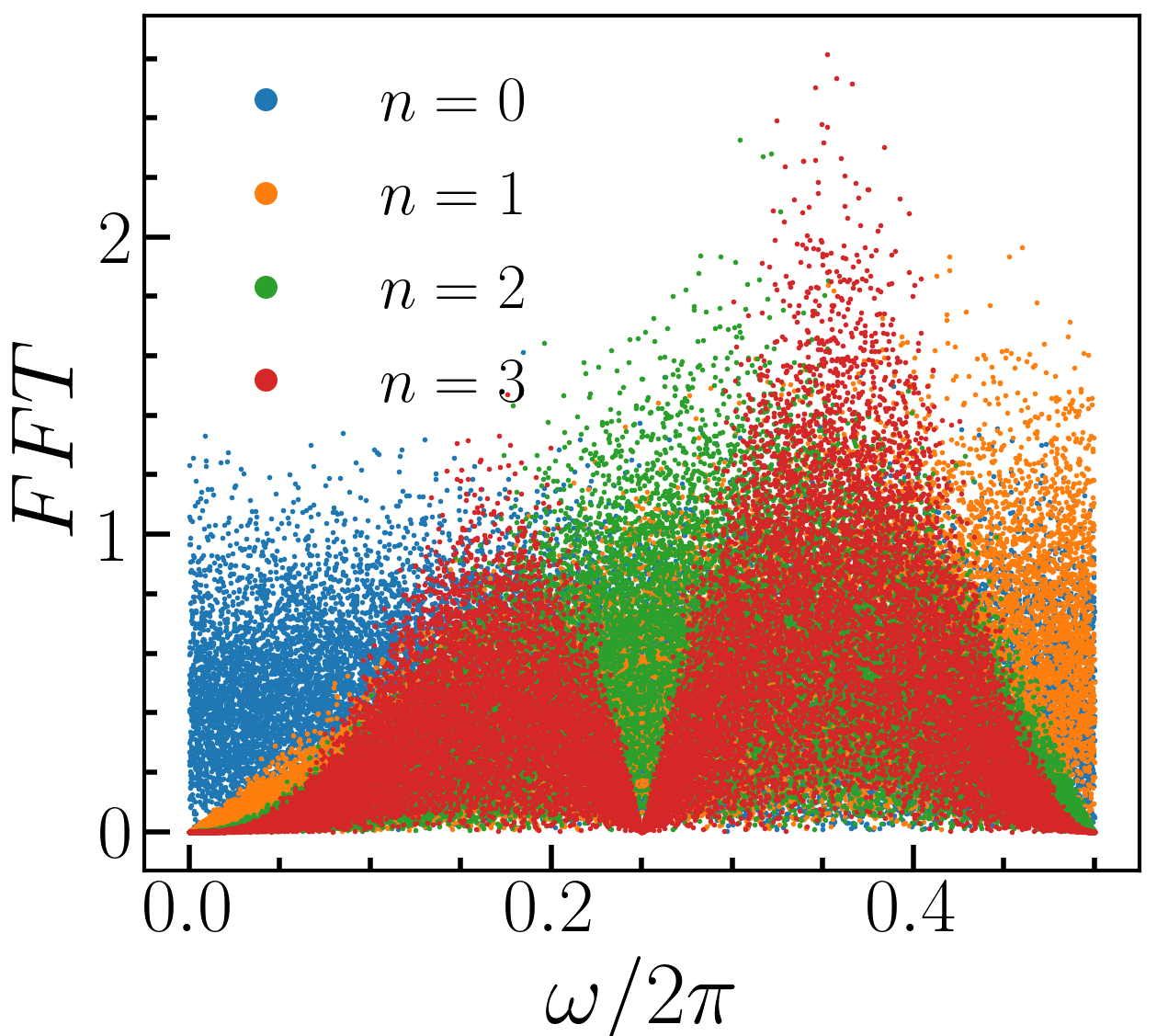}
	\includegraphics[width=0.5\linewidth]{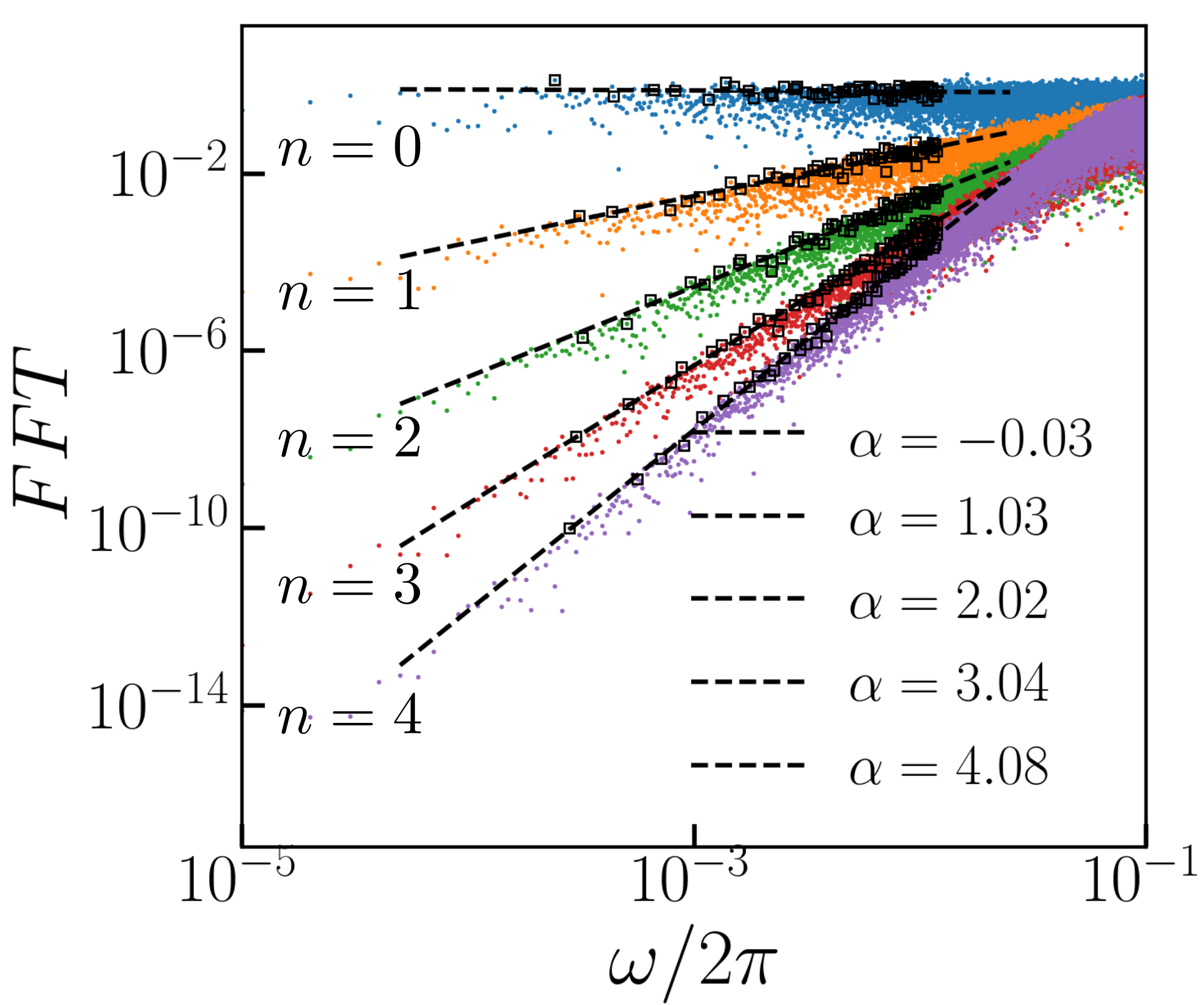}
	\caption{Left: Fourier spectrum of random multipolar sequences for different $n$. Right: Log-Log plot for the spectrum. Suppression at low frequencies scales algebraically as $\omega^n$.  }
	\label{fig:fft}
\end{figure}
As seen in the left panel of Fig.~\ref{fig:fft}, for a random sequence (blue, $n=0$), the spectrum fluctuates randomly over all frequencies with a flat envelop. Once imposing the multipolar structure, several suppression of frequencies appears at different positions, for example at $\omega=0$ for $n\ge 1$, and ar $\omega=\pi$ for $n\ge 2$.  By plotting the spectrum on a  Log-Log scale in the right panel, we can fit the envelope of the spectrum and identify the scaling of suppression at low frequencies as $\omega^{n}$. 

\begin{figure}[h]
	\centering
	\includegraphics[width=0.51\linewidth]{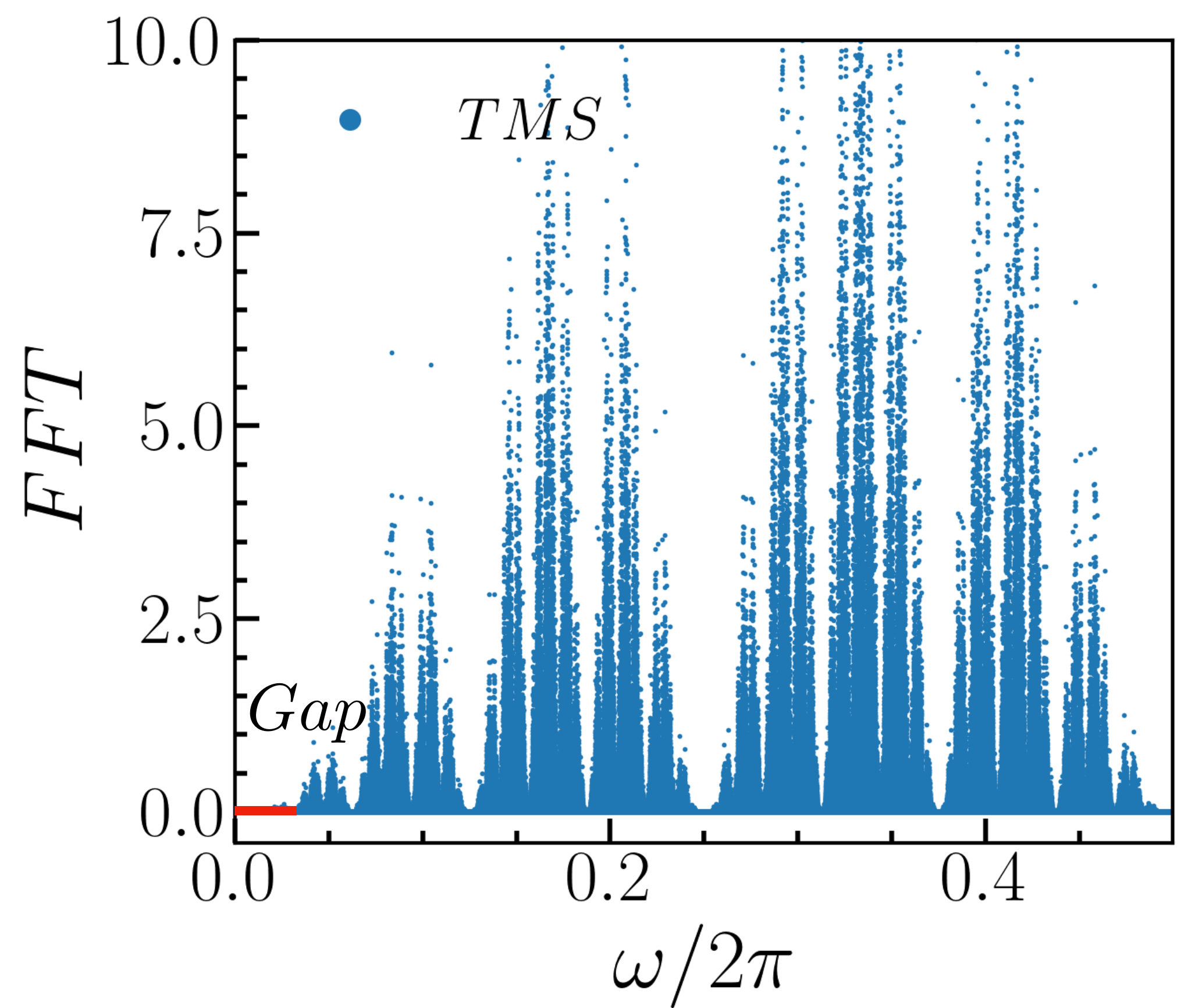}
	\caption{Frequency spectrum for the TMS.}
	\label{fig:fft_TMS}
\end{figure}
The quasiperiodic Thue-Morse sequence (TMS) corresponds to the random multipolar sequence in the $n\to \infty$ limit, and practically we use $n=18$ to approximate it. In this limit, the low frequency spectrum scales as $\omega^{n\to \infty}$ thus fast approaching zero for $\omega\to0$. Hence, the algebraic scaling converts to a gap at low frequency as observed in Fig.~\ref{fig:fft_TMS}.

\paragraph{Analytical derivation of the low frequency behavior.--} The low frequency scaling can be rationalized by the  autocorrelation function of the sequence. The autocorrelation function
for the $n-$multipolar sequence $x^{(n)}(t)$ is defined as
\begin{eqnarray}
R^{(n)}(t,\tau) = \langle x^{(n)}(t) x^{(n)}(t+\tau)\rangle,
\end{eqnarray}
where $\langle \dots\rangle$ denotes the average over both different random realization and $t$. In this case, $R^{(n)}(t,\tau)$ becomes translation invariant in $t$. 
For the fully random sequence $n=0$, one obtains the standard white noise form  $R^{(0)}(\tau)=\delta_{\tau,0}$;  for the dipolar sequence $n=1$, $R^{(1)}(\tau)=\delta_{\tau,0}+\delta_{\tau,0}-\delta_{\tau,1}-\delta_{\tau,-1}$ in each dipolar unit cell; etc.
The Fourier transformation $\hat{R}^{(n)}(\omega)$, which is also known as the spectral density, can be defined as
\begin{eqnarray}
\hat{R}^{(n)}(\omega) =  \int_{-\infty}^{\infty}d\tau R^{(n)}(\tau) e^{i\omega \tau}.
\end{eqnarray}
The dipolar sequence has $\hat{R}^{(1)}(\omega)=2-2\cos(\omega)$, exhibiting the scaling $\omega^2$ at low frequencies. 

For $n>1$, one can derive the scaling in the following way.
First, we know the $n-$multipolar unit cells are formed by two anti-aligned $(n-1)-$multipoles of size $2^{n-1}$. Accordingly, each $(n-1)-$multipole contributes $R^{(n-1)}(\tau)$ to the anticorrelation, and the interplay between different types of $(n-1)-$multipoles gives the negative contribution $-R^{(n-1)}(\tau\pm2^{n-1})$, where $\pm2^{n-1}$ is introduced due to the relative displacement of the two $(n-1)-$multipoles. Therefore we arrive at the following iterative relation
\begin{eqnarray}
\begin{aligned}
R^{(n)}(\tau) &= 2R^{(n-1)}(\tau)
\\&-R^{(n-1)}(\tau-2^{n-1})-R^{(n-1)}(\tau+2^{n-1}),
\end{aligned}
\end{eqnarray}
corresponding to the following relation for the spectral density
\begin{eqnarray}
\hat{R}^{(n)}(\omega) =  \hat{R}^{(n-1)}(\omega) \left[2-2\cos(2^{n-1} \omega)\right],
\end{eqnarray}
with the initial condition $ \hat{R}^{(0)}(\omega) =1.$ In the end, we arrive at
\begin{eqnarray}
\hat{R}^{(n)}(\omega)  = 2^n\prod_{j=1}^{n} \left[ 1-\cos(2^{j-1}\omega)\right],
\end{eqnarray}
which has multiple zero points depending on the tunable order $n$. In particular, one zero point locates at $\omega=0$, and the nearby suppression scales as $\omega^{2n}$. 
The following relation~\cite{grimmett2001probability}
\begin{eqnarray}
\hat{R}^{(n)}(\omega) = \lim\limits_{T\to \infty}\left\langle |\hat{x}^{(n)}(\omega)|^2\right\rangle,
\end{eqnarray}
now helps us to identify the scaling behavior of the Fourier transformation of $x^{(n)}(t)$ defined as
\begin{equation}
\hat{x}^{(n)}(\omega)=\frac{1}{\sqrt{T}} \int_{0}^{T} x^{(n)}(t) e^{-i \omega t} d t.
\end{equation}
From Cauchy-Schwarz inequality, we have
\begin{eqnarray}
\label{eq.average}
\left\langle |\hat{x}^{(n)}(\omega)|\right\rangle \leq \sqrt{\left\langle |\hat{x}^{(n)}(\omega)|^2\right\rangle},
\end{eqnarray}
suggesting that $\left\langle |\hat{x}^{(n)}(\omega)|\right\rangle $ is upper bounded by $\sqrt{\hat{R}^{(n)}(\omega)}\propto \omega^n$ for a $n-$multipolar sequence given a long enough time window of integration $T\to \infty$. Although Eq.~\ref{eq.average} is valid for the average of the Fourier spectrum, we also expect it to impose the same bound $\omega^n$ for a single sequence realization in accordance with the numerical results of Fig.~\ref{fig:fft}. 

\section{Fermi's golden rule(FGR)}
Here we discuss FGR in detail. As introduced in the main content, we consider the periodically driven system described by the Hamiltonian
\begin{eqnarray}
\hat{H}(t)=\hat{H}_{0}+g(t) \hat{K},
\end{eqnarray}
where $g(t) \hat{K}$ is a weak time-periodic perturbation and can be decomposed as
\begin{equation}
\label{eq.decompose}
g(t) \hat{K}=\sum_{m>0} 2g_{m} \sin (m \Omega t) \hat{K}.
\end{equation}
After a short initial transient dynamics, in the linear response regime, the system absorbs energy independently from each Fourier mode $m$~\cite{Rigol2019heating}. 
The average rate of energy absorption over a cycle is
\begin{equation}
\label{eq.rateofenergy}
\dot{E}(t)=\sum_{m>0} \dot{E}_{m}(t),
\end{equation} 
where $\dot{E}_{m}(t)$ is expected from Fermi's golden rule as \cite{Rigol2019heating}
\begin{equation}
\label{eq.defenergy}
\begin{array}{c}
\dot{E}_{m}(t)=2 \pi g_{m}^{2} \sum_{i, f}\left|\left\langle E_{f}^{0}|\hat{K}| E_{i}^{0}\right\rangle\right|^{2}\left(E_{f}^{0}-E_{i}^{0}\right) P_{i}^{0}(t) \\
\times \delta\left(E_{f}^{0}-E_{i}^{0} \pm m \Omega\right),
\end{array}
\end{equation}
where $| E_{i}^{0}\rangle$ are the eigenstates of the static $\hat{H}_0$, and 
$
P_{i}^{0}(t)=\left\langle E_{i}^{0}|\hat{\rho}(t)| E_{i}^{0}\right\rangle,
$ with the density matrix $\hat{\rho}(t).$ The thermalization rate is defined 
as \begin{eqnarray}\label{eq.rate}
\Gamma(t)=\sum_{m>0} \Gamma_{m}(t),
\end{eqnarray}
with $ \Gamma_{m}(t)=\dot{E}_{m}(t) /\left[E_{\infty}-E(t)\right]$ and $E_{\infty}$ is the energy at infinite temperature, which turns to be zero in our case. $\Gamma_{m}(t)$ remains to be nearly constant~\cite{Rigol2019heating}, and the inverse of the rate $1/\Gamma$ can be used to approximate the thermalization time. 
Thus, the heating rate scales exponentially with driving frequency in the fast driving regime as
\cite{abanin2015exponentially,ho2018bounds,Mori2016}
\begin{equation}
\label{eq.scaling_gamma}
\begin{aligned}
\Gamma_{m}=g_m^2 Ae^{-m\Omega/\epsilon},
\end{aligned}
\end{equation}
consequently the thermalization time scales as $e^{m\Omega}$ for Floquet systems.
There are two undetermined system-dependent parameters $A,\epsilon$ (the latter is interpreted as the effective local band width~\cite{machado2019}),  and both of them are independent of the frequency $m\Omega$. Eq.~\ref{eq.scaling_gamma} also implies the scaling versus driving amplitude as $g_m^2$, which is crucial for our following discussion.

For $n-$multipolar driving, instead of the discrete Fourier decomposition for periodic driving, the driving has a continuous spectrum, and we use the following ansatz to mimic the driving
\begin{equation}
\label{eq.decompose_new}
g(\tau) \hat{K}=\int dx g_x \sin (x \Omega \tau) \hat{K},
\end{equation}
where $g_x$ is the amplitude for a continuous variable $x$. In particular, we are interested in the function $g_x=x^n$ depicting the envelope of the $n-$multipolar random sequence at low frequencies. 

Again by assuming that the system absorbs energy from each mode independently, the heating rate turns into an integral as
$
\label{eq.rate_integral}
\Gamma=\int_0^{\infty} dx \Gamma_x,$
with
$
\Gamma_{x}=g_x^2 Ae^{-x\Omega/\epsilon}.
$
By inserting $g_x=x^n$ and only focusing on the scaling behavior of frequency, one can analytically solve the integral as
\begin{eqnarray}
\label{eq.int}
\begin{aligned}
\Gamma&\propto\int_0^{\infty} dx x^{2n} e^{-x\Omega}=\int_0^{\infty} d(x\Omega)(x\Omega)^{2n} e^{-x\Omega} \Omega^{-2n-1}\\
&=\Omega^{-2n-1}\int_0^{\infty} dy y^{2n} e^{-y}=\Omega^{-2n-1}(2n)!,
\end{aligned}
\end{eqnarray}
where we use the formula~\cite{abramowitz1948handbook}
\begin{equation}
\begin{aligned}
\int y^{n} e^{c y} d y =e^{c y} \sum_{i=0}^{n}(-1)^{n-i} \frac{n !}{i ! c^{n-i+1}} y^{i}.
\end{aligned}
\end{equation}
The inverse of $\Gamma$ gives the thermalization time scaling $(1/T)^{2n+1}$ in accordance with the numerical results presented in Fig.~\ref{fig:multipole}.

The TMS generates a  gap around $\omega=0$ (Fig.~\ref{fig:fft_TMS}). Suppose the gap is larger than the local band width, one can approximately define the function $g_x=\delta_{x,x_0}$ where $x_0\Omega$ denotes the size of the gap. In this case, the heating rate reduces to that of the Floquet systems, hence an exponential scaling of the thermalization time is expected. Energy absorption of single modes with energy larger than $x_0\Omega$ will be even more suppressed, resulting in a thermalization time later than $e^{x_0\Omega}$. 

\section{Initial state dependence of prethermalization}
\begin{figure}
	\centering
	\includegraphics[width=0.99\linewidth]{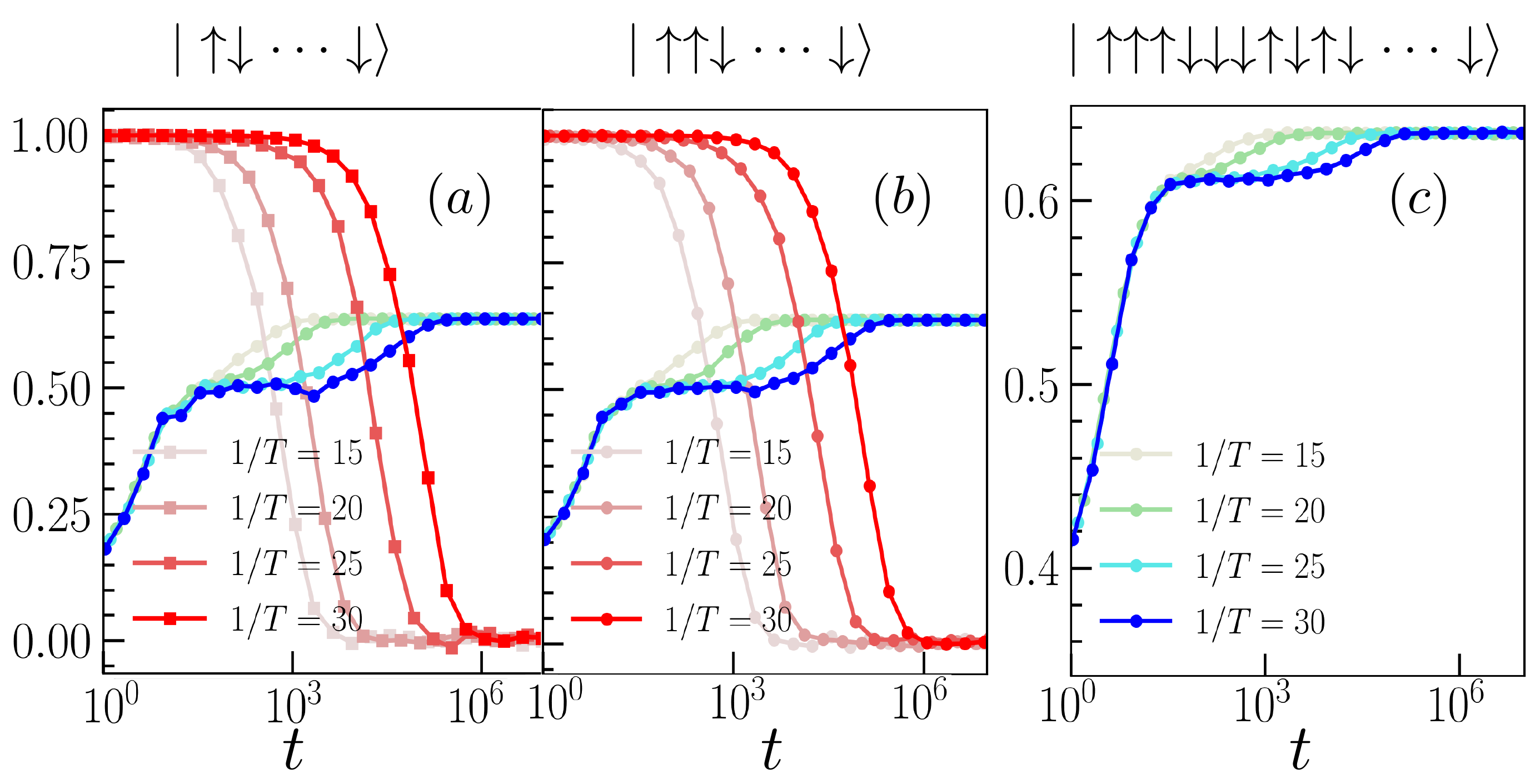}
	\caption{Dynamics of entanglement entropy(blue) and energy (red) for different initial states, with the same parameters as in Fig.~\ref{fig:dynamics}.}
	\label{fig:initialstate}
\end{figure}

\begin{figure}
	\centering
	\includegraphics[width=0.89\linewidth]{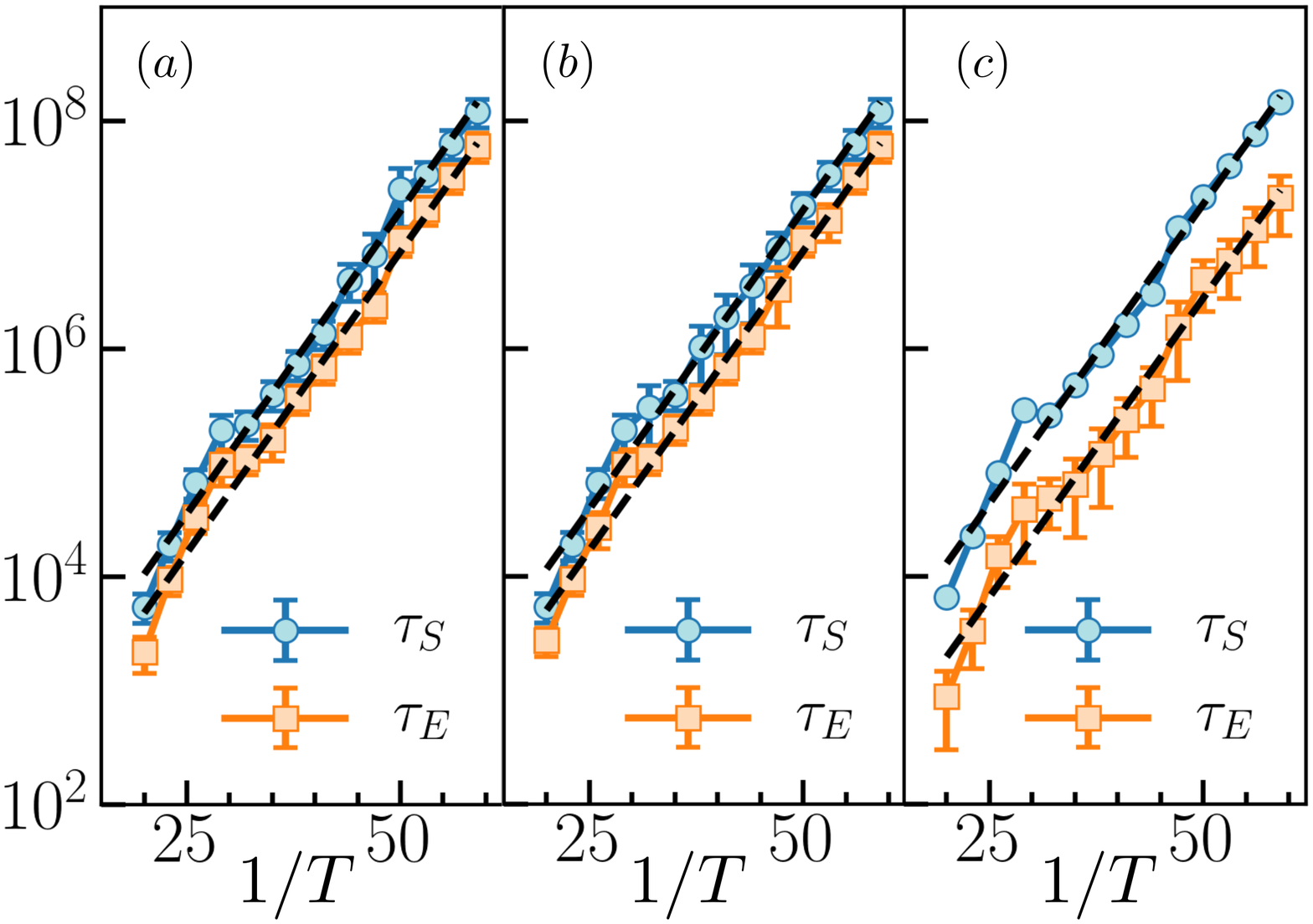}
	\caption{Thermalization time scaling versus $T^{-1}$. Note, thresholds for entanglement and energy are chosen as $S_L/(L/2)\sim 0.58\pm0.02$ and $\langle H_{F}^0\rangle_t/\langle H_{F}^0\rangle_0\sim 0.7\pm0.1$ for panel (a)(b), and $\langle H_{F}^0\rangle_t/\langle H_{F}^0\rangle_0\sim 0.85\pm0.1,$ $S_L/(L/2)\sim 0.63$ for panel (c).  }
	\label{fig:scalingtest}
\end{figure}
In the main content, the dynamics induced by Thue-Morse drives use all spins pointing down as the initial state. Here, we compare the dynamics for different initial states deviating from the fully polarized state with various numbers of domain walls. The results are plotted in Fig.~\ref{fig:initialstate} where we confirm the lack of energy absorption and a prethermal plateau as long as the initial state does not deviate too much from the ferromagnetic state.

The thermalization time are plotted in Fig.~\ref{fig:scalingtest} on a Log scale, where the exponential dependence on $1/T$ can be clearly seen. The slope of the exponential fitting is insensitive to threshold values. As seen in the panel (c), different thresholds are applied to estimate $\tau_{S/E}$ while the slope of the fitting remain the same.

\section{Finite size effect}
\begin{figure}[h]
	\centering
	\includegraphics[width=0.99\linewidth]{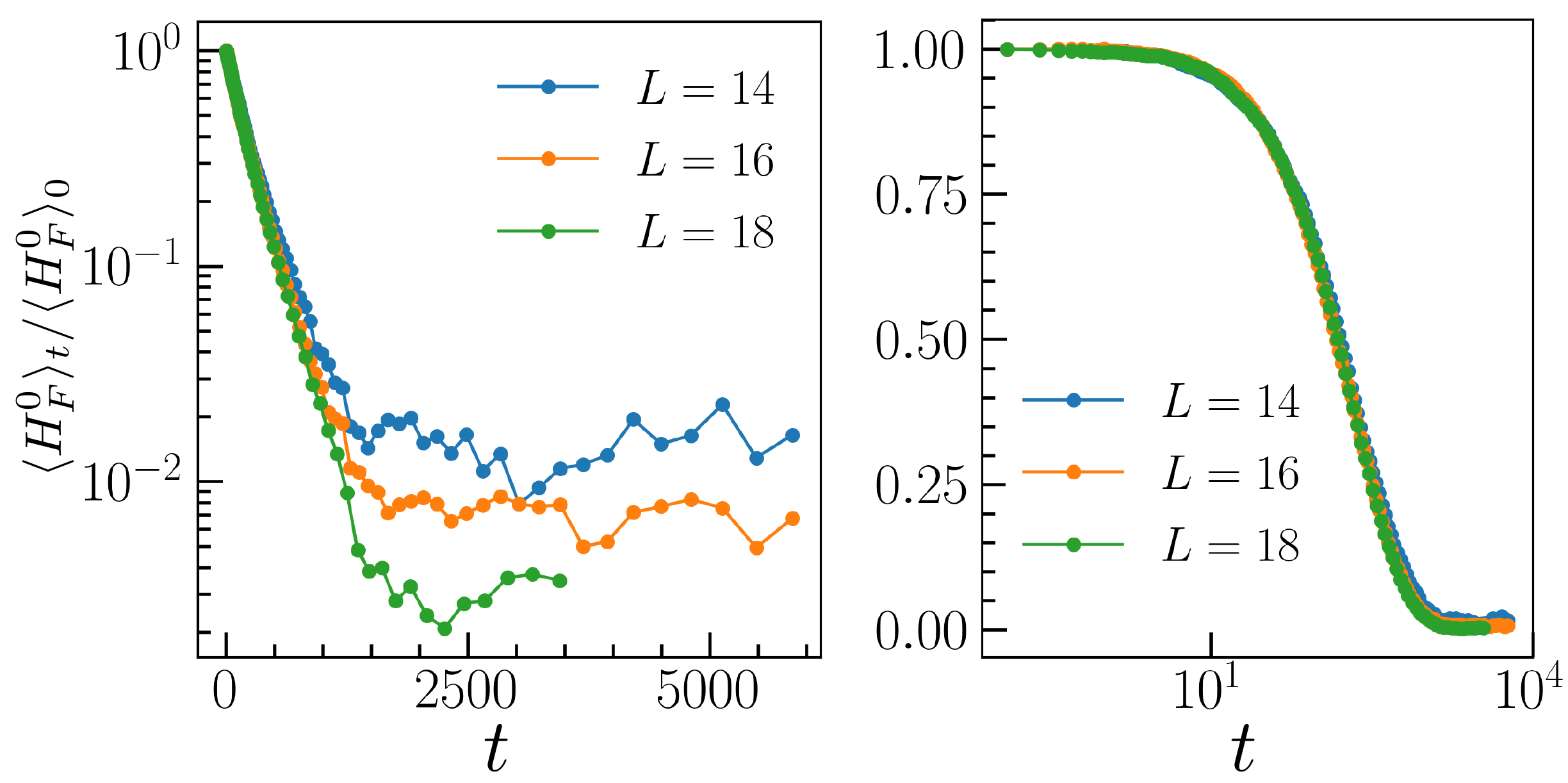}
	\caption{Energy dynamics for different system sizes induced by a random dipolar drives, using parameters  $\{J_z,J_x,B_x,B_z,B_0,T^{-1}\}=\{1,0.243,0.809,0.357,0.21,16\}$. Finite size effect is negligible for $L\geq 16$ before the final relaxation, where larger system size results in a lower energy value approaching to zero.}
	\label{fig:sizescaling}
\end{figure}

Here we compare the dynamics obtained for different system sizes. As seen in the right panel in Fig.~\ref{fig:sizescaling} where the time is plotted on a log scale, one can hardly tell the differences between three results.  In the left panel where the energy is plotted on a log scale, for $t<1000,$ the results show exponential decay and converge well for $L\geq 16$. In particular, at the mean energy used to determine thermalization time scaling in Fig.~\ref{fig:multipole}, $\langle H_{F}^0\rangle_t/\langle H_{F}^0\rangle_0\sim 0.95\pm0.01$, no finite size effect can be observed. 

Finite size effect only becomes visible at late times $t>1500$ not used for obtaining the scaling relation. There, the mean energy decreases notably for increasing system sizes, which is expected to be zero in the thermodynamic limit.

\section{Comparison between $n=0$ and $1$-RMD for weak driving} 
\begin{figure}[h]
	\centering
	\includegraphics[width=0.655\linewidth]{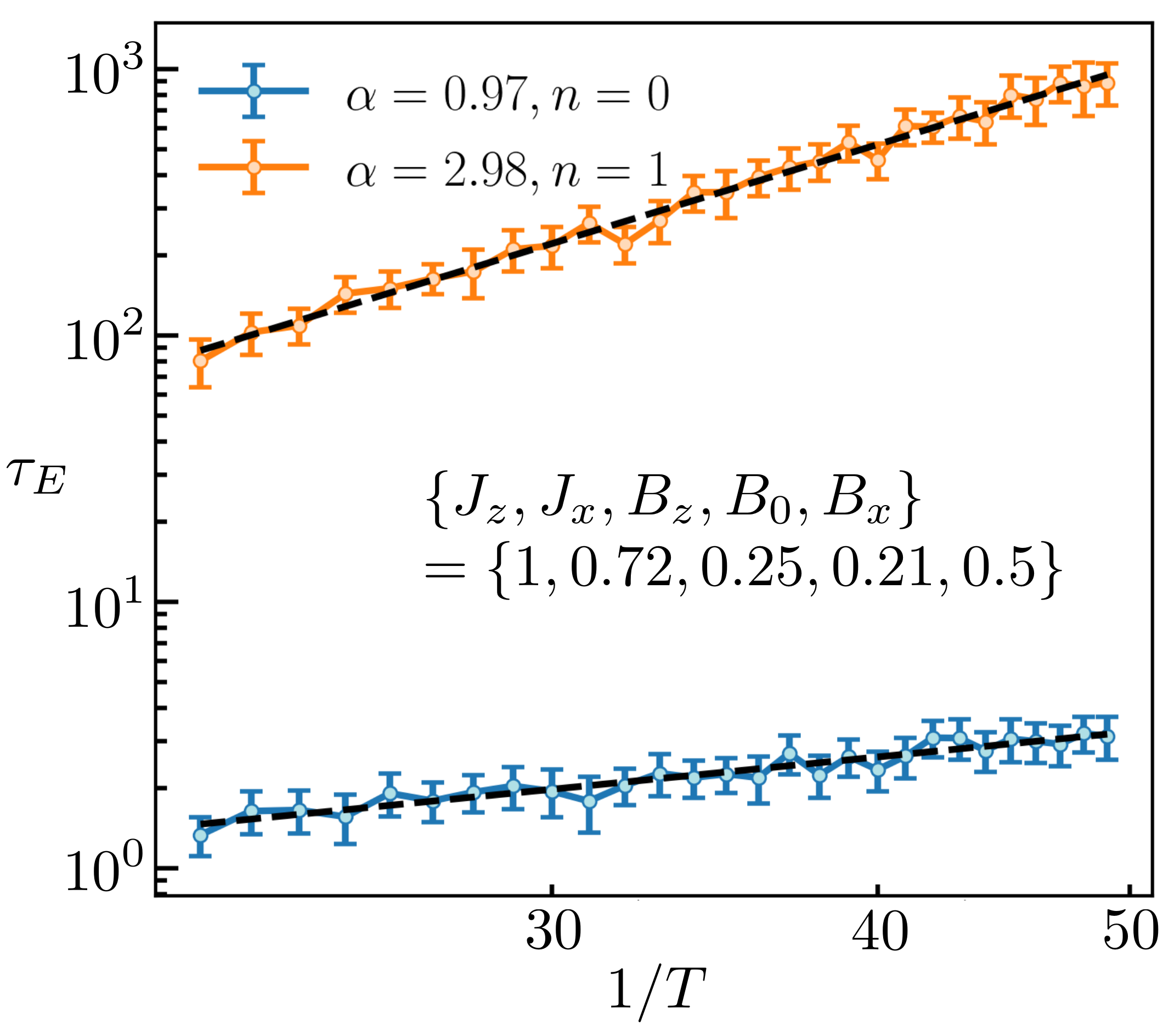}
	\caption{Thermalization time scaling for $n=0$ and 1-RMD in the weak driving regime $B_x=0.5$. The algebraic scaling $T^{-2n-1}$ is valid in both cases.}
	\label{fig:RandomParm5}
\end{figure}

{In Fig.~\ref{fig:RandomParm5}, we show that  for weak driving amplitude $B_x=0.5$, the algebraic dependence $\tau_E\sim T^{-2n-1}$ works for both $n=0$ and $n=1$. A transition from $\tau_E\sim T^{-1}$ to $\tau_E\sim$ const happens by increasing $B_x$. As shown in Fig.~\ref{fig:multipole} for large driving amplitudes $B_x=3.2$, the algebraic scaling $T^{-2n-1}$ of thermalization time is only valid for $n=1$.}

\section{Prethermal DTC with imperfect rotation}
\begin{figure}[h]
	\centering
	\includegraphics[width=0.65\linewidth]{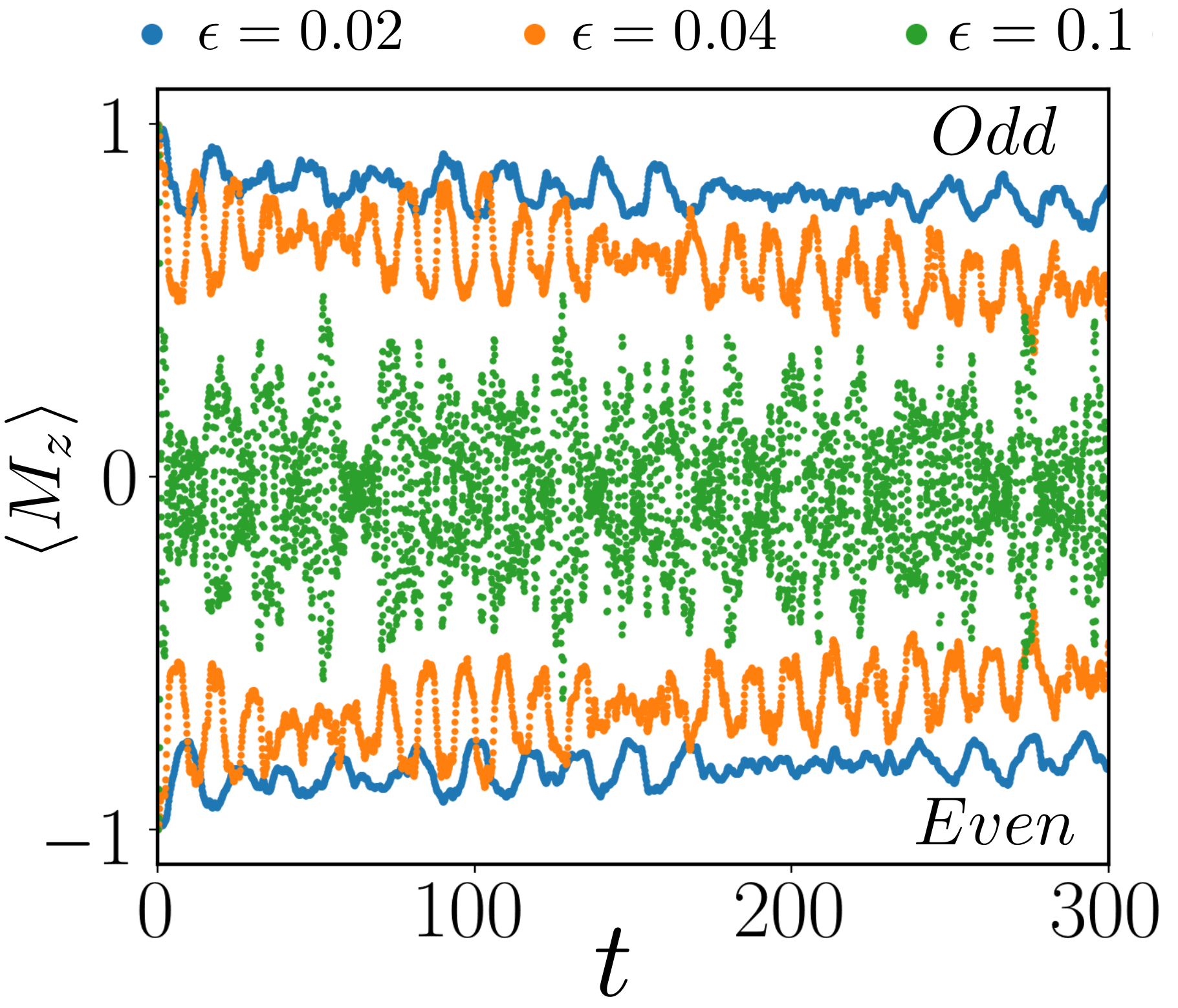}
	\caption{Dynamics of DTC with imperfect rotation. 
		$\{J_z,J_x,B_z,B_0,B_x\}= \{1,0.315,0.21,-0.05,0.25\}, L=18,T^{-1}=30$ }
	\label{fig:pert_DTC}
\end{figure}

Here we discuss the stability of the DTC with respect to imperfect spin flips
$
X = \exp \left(i \frac{\pi+\epsilon}{2} \sum_{i} \sigma_{i}^{x}\right)
$
used in Eq.~\ref{eq.DTC}
for a small $\epsilon$. The magnetization dynamics for different perturbations $\epsilon$ are plotted in Fig.~\ref{fig:pert_DTC} (driving rate is $T^{-1}=30$). 
The prethermal DTC remains stable for small perturbations, for instance $\epsilon=0.02$.
For $\epsilon= 0.04$, the amplitude of the period-doubling oscillation exhibits notable decay. For $\epsilon=0.1$, no time-crystalline order exists and the oscillations around zero magnetization are induced by finite size effect.
It is not very surprising to see the phase is fragile to rotation imperfections as there is no mechanism to protect the system from thermalization. It will be interesting to investigate if disorder, long-range interactions or dissipation are capable of stabilizing the prethermal DTC further.

\end{document}